\documentclass[journal=jacscomm,manuscript=communication]{achemso}
\usepackage[version=3]{mhchem} % Formula subscripts using \ce{}
\usepackage[T1]{fontenc}       % Use modern font encodings
\usepackage{longtable}
\usepackage{multirow}
\usepackage{array}
\usepackage[english]{babel}
\usepackage{booktabs} 
\usepackage{amsmath}
\usepackage{graphicx}
\usepackage{subfig}
\author{Srimanta Pakhira}
\altaffiliation{Condensed Matter Theory, National High Magnetic Field Laboratory (NHMFL), Scientific Computing Department, Materials Science and Engineering, Florida State University, Tallahassee, Florida, 32310, USA.}
\author{Kevin P. Lucht}
\altaffiliation{Condensed Matter Theory, National High Magnetic Field Laboratory (NHMFL), Scientific Computing Department, Materials Science and Engineering, Florida State University, Tallahassee, Florida, 32310, USA.}
\author{Jose L. Mendoza-Cortes}
\email{mendoza@eng.famu.fsu.edu}
\phone{+1-850-410-6298}
\fax{+1-850-410-6150}
\altaffiliation{Condensed Matter Theory, National High Magnetic Field Laboratory (NHMFL), Scientific Computing Department, Materials Science and Engineering, Florida State University, Tallahassee, Florida, 32310, USA.}
\affiliation{Department of Chemical \& Biomedical Engineering, FAMU-FSU Joint College of Engineering, and High Performance Materials Institute (HPMI), Florida State University, Tallahassee, Florida, 32310, USA.}

\title {Iron Intercalated Covalent-Organic Frameworks: First Crystalline Porous Thermoelectric Materials} 
\begin{document}

%\begin{tocentry}
%
%Some journals require a graphical entry for the Table of Contents.
%This should be laid out ``print ready'' so that the sizing of the
%Inside the \texttt{tocentry} environment, the font used is Helvetica
%8\,pt, as required by \emph{Journal of the American Chemical
%The surrounding frame is 9\,cm by 3.5\,cm, which is the maximum
%permitted for  \emph{Journal of the American Chemical Society}
%graphical table of content entries. The box will not resize if the
%content is too big: instead it will overflow the edge of the box.
%
%This box and the associated title will always be printed on a
%separate page at the end of the document.
%
%\end{tocentry}

\begin{abstract}
  Covalent-organic frameworks (COFs) are intriguing platforms for designing functional molecular materials. Here, we present a computational study based on van der Waals dispersion-corrected hybrid density functional theory calculations to analyze the material properties of boroxine-linked and triazine-linked intercalated-COFs. The effect of Fe atoms on the electronic band structures near the Fermi energy level of the intercalated-COFs have been investigated. The density of states (DOSs) computations have been performed to analyze the material properties of these kind of intercalated-COFs. We predict that COFs's electronic properties can be fine tuned by adding Fe atoms between two organic layers in their structures.  The new COFs are predicted to be thermoelectric materials. These intercalated-COFs provide a new  strategy to create thermoelectric materials within a rigid porous network in a highly controlled and predictable manner.
  
\end{abstract}

\section{Introduction}
Recently, as a new class of emerging hybrid porous materials, covalent-organic frameworks (COFs) are an attractive class of crystalline porous materials that can integrate organic units with atomic precision into periodic structures \cite{El-Kaderi2007}. COFs constitute a class of synthetic materials obtained through the ordered polymerization of organic monomers and form a crystalline framework.  The COF materials structures contain generally light elements, such as H, B, C, N, and O. COFs combine the thermodynamic strength of covalent bonds as those found in diamonds and boron carbides, with the functionality of organic units \cite{Cote2005}. Thus they have high structural regularity and diversity, easy modification of frameworks, high porosity and structural flexibility which make them an ideal candidate for various potential applications in fields of science particularly gas adsorption, gas storage/uptake (H$_2$, CH$_4$, etc.) \cite{Mendoza-Cortes2012, Mendoza-Cortes2010, Pramudya}, including energy applications \cite{Furukawa} from fuel cells to supercapacitors \cite{SooHan2009, Mendoza-Cortes2012a}.   
    
Present computational and theoretical research shows many interesting electronic properties of two-dimensional (2D) COF materials with honeycomb-like structures (like graphene) \cite{Han2008}, such as: half-metallicity, low intrinsic conductivity, semiconducting, and photoconductivity \cite{Zhang2013, Ding2011}. In addition, because of their $\pi$-stacked layered structures, which can drive the electronic interactions between the nearest neighbor sheets, new properties are expected to emerge from the interplay between the properties of the electronically coupled individual frameworks.

To the best of our best knowledge, the fundamental research into the bulk pristine COFs and intercalated-COFs have not yet been extensively studied, thus the works on intrinsic properties of the bulk pristine COFs and intercalated-COFs still remains an open opportunity for material science. In the present work, we have studied the structures and material properties of the bulk boroxine-linked and triazine-linked -COF materials using first-principle calculations. Here we have shown how the material properties are altered by the addition of Fe atoms between two organic layers in COFs. We also address how the metal atoms affects the band structures and DOSs of the COFs. In this study, three different types of boroxine-linked and triazine-linked intercalated-COF materials are taken into consideration, which are: COF-SU-01 (Fe$_3$-COF i.e. three Fe atoms were intercalated in the pristine COF per unit cell), COF-SU-02 (Fe$_4$-COF i.e. four Fe atoms were intercalated in the pristine COF per unit cell), and COF-SU-03 (Fe$_5$-COF i.e. five Fe atoms were intercalated in the pristine COF per unit cell) along with the pristine COF as shown in Figure \ref{fig:Optimized_Geometry}. Thus this work presents the first comprehensive investigation of bulk boroxine-linked and triazine-linked COF materials with Fe intercalated. These intercalated-COF materials may have interesting thermoelectric and electronic applications.   
 
First-principle calculations based on hybrid density functional theory (DFT) as implemented in CRYSTAL14 \cite{Dovesi2014} was used to perform all the computations. The geometries of bulk crystal structure of the aforementioned pristine COF and intercalated-COF materials were optimized by the dispersion-corrected B3LYP hybrid DFT \cite{Becke1993, Becke1988, Lee1988, Pakhira2015, Pakhira2016}, i.e. B3LYP-D2 \cite{Grimme2006, Pakhira2012, Pakhira2013, Pakhira2015a} which has been shown to give correct electronic properties of the 2D/3D materials \cite{Pakhira2016b, Lucht2015}. The semi-empirical Grimme-D2 dispersion corrections were added in the present calculations in order to incorporate van der Waals (vdW) dispersion effects on the system.\cite{Grimme2006, Pakhira2012, Pakhira2013} In the present computation, triple-zeta valence with polarization quality (TZVP) Gaussian basis sets were used for H, B, C, N, O and Fe atoms \cite{Peintinger2013}. Integration inside of the first Brillouin zone were sampled on 4x4x16 Monkhorst-Pack \cite{Monkhorst1976} k-mesh grids for all the intercalated-COFs and pristine COF materials for the structure optimization, and on 20x20x20 Monkhorst-Pack k-mesh grids for the calculations of band structure and density of states (DOSs). Visualization and analysis were performed using VESTA \cite{Momma2011}. 

\section{Results and discussion}

\subsection{Outline}

\begin{figure}
	\includegraphics[width=0.99\linewidth]{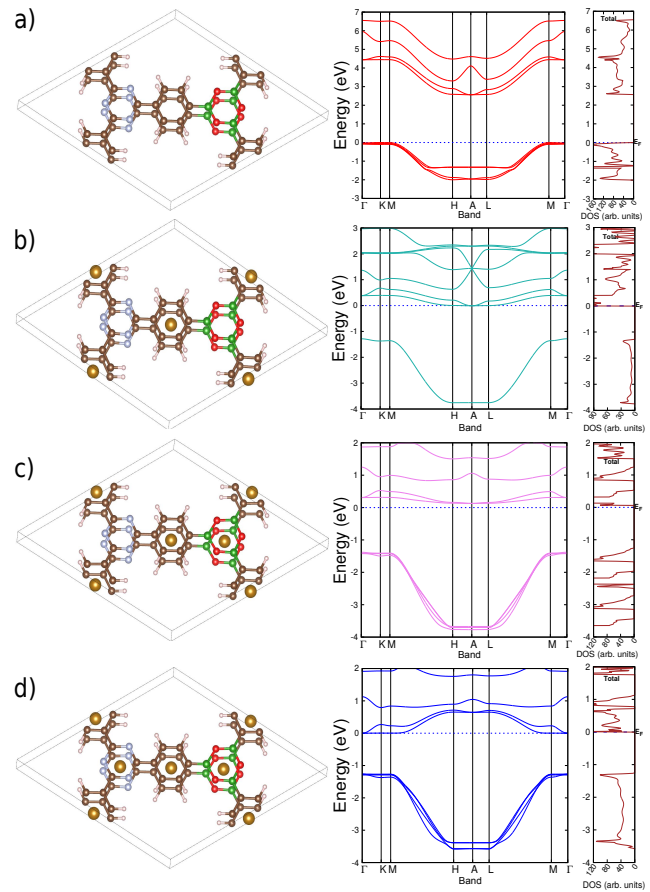}
	\caption{Optimized crystal structures, band structure and DOSs of a) COF-SU-00, b) COF-SU-01, c) COF-SU-02, and  d) COF-SU-03 materials.}
	\label{fig:Optimized_Geometry}
	%	\vspace{0mm}
\end{figure}

The optimized crystal structures of the pristine and intercalated-COFs, COF-SU-00, COF-SU-01, COF-SU-02, and COF-SU-03, are shown in Figure \ref{fig:Optimized_Geometry}a, \ref{fig:Optimized_Geometry}b, \ref{fig:Optimized_Geometry}c, and \ref{fig:Optimized_Geometry}d, respectively. Fe atoms have been placed at the centroid of three benzene rings between two layers in one unit cell of COF-SU-01. The fourth Fe atom is placed at the centroid of boroxine ring in COF-SU-01 to form COF-SU-02, and the fifth Fe atom is placed at the centroid of triazine ring in COF-SU-02 to form COF-SU-03. The coordinates of optimized crystal structures of the COFs are reported in the Supporting Information.  The COFs, presented here, are ideal honeycomb-like structures with one triazine, one boroxine, and five benzene rings in one unit cell with $p-6m2$ symmetry in their crystal structures. Due to the similar crystal structure but slight different chemistry, it is interesting to compare their basic electronic structures and properties. The equilibrium lattice constants ($a$, $b$, $c$), the distance between two layers ($d$), various average bond distances (C-C, C-O, O-B, B-C, C-N) and band gap ($E_g)$ of the pristine and intercalated-COFs are reported in Table \ref{tbl:Table1}. Both the lattice constants and various bond distances gradually changed when the Fe atoms were added in the COFs. In all the intercalated-COFs studied here, the boroxine and triazine rings are slightly different from the benzene ring. As for example, both the boroxine and triazine rings are not a perfect hexagon in the COF-SU-01, and the angles (\angle{BOB} and \angle{NCN}) are 121.6$^{\circ}$ and 124$^{\circ}$ respectively. This is because the electronegativity of the O and N atoms is stronger (i.e. there is a lone pair of electrons), and both the boroxine and triazine rings have very little aromatic character compared to pure benzene rings. The distances $d$ between two identical layers in the COF-SU-00, COF-SU-01, COF-SU-02, and COF-SU-03 materials were increased as more Fe atoms were added between the layers (see Table \ref{tbl:Table1}). \\

%\begin{table}
%	\caption{The lattice constants (a, b, c) and average bond distances (in \AA) of the pristine COF and iron intercalated-COF, COF-SU-01, COF-SU-02, and COF-SU-03, materials.
%	\label{tbl:Table1}
%	\tabcolsep=4.0pt
%	\begin{tabular}{lccccccccr}
%		\toprule
%		COFs         &a     &b     & c    & C-C  & C-B  & B-O  & C-N & d & $E_g$ \\
%		\midrule
%		Pristine COF &14.85 &14.85 &3.31  & 1.39 & 1.53 & 1.38 & 1.34 & 3.241 & 2.6 \\
%		COF-SU-01    &14.83 &14.83 & 3.44 & 1.44 & 1.51 & 1.39 & 1.34 & 3.443 & 1.2 \\
%		COF-SU-02    &14.95 &14.95 & 3.45 & 1.44 & 1.50 & 1.43 & 1.34 & 3.448 & 1.4 \\
%		COF-SU-03    &15.03 &15.03 & 3.45 & 1.43 & 1.51 & 1.43 & 1.39 & 3.440 & 1.4 \\
%		\bottomrule
%	\end{tabular}
%\end{table}

\begin{table}
	\vspace{0mm}
	\caption{The lattice constants ($a$, $b$, $c$), average bond distances and intercalation distance ($d$) between two layers of the COFs and band gap energy $E_g$ of the pristine COF and iron intercalated-COF, COF-SU-01, COF-SU-02, and COF-SU-03, materials. The band gap energy and the bond distances are expressed in eV and \AA, respectively.}
	\label{tbl:Table1}
	\tabcolsep=1.50pt
	\begin{tabular}{lccccccccr}
		\toprule
		COFs         & $a$  & $b$  & $c$  & C-C  & C-B  & B-O  & C-N  & $d$     & $E_g$ \\
		\midrule
		COF-SU-00    &14.85 &14.85 &3.31  & 1.39 & 1.53 & 1.38 & 1.34 & 3.241 & 2.60 \\
		COF-SU-01    &14.83 &14.83 & 3.44 & 1.44 & 1.51 & 1.39 & 1.34 & 3.443 & 1.20 \\
		COF-SU-02    &14.95 &14.95 & 3.45 & 1.44 & 1.50 & 1.43 & 1.34 & 3.448 & 1.38 \\
		COF-SU-03    &15.03 &15.03 & 3.45 & 1.43 & 1.51 & 1.43 & 1.39 & 3.440 & 1.18 \\
		\bottomrule
	\end{tabular}
%	\vspace{-5mm}
\end{table}

%\begin{figure}
%	\vspace{-15mm}
%	%  and \texttt{figure}, the class also recognises\\
%	%  \texttt{scheme}, \texttt{chart} and \texttt{graph}.
%	\includegraphics[width=0.99\linewidth]{./FIG1}
%	\caption{Band structure and DOSs of the Fe3-COF material.}
%	\label{BG_Fe3_Triazene_BAND-DOSS_NORM}
%%	\vspace{-15mm}
%\end{figure}

\subsection{Results and discussions}
After obtaining the optimized geometry of the unit cell for the pristine COF and intercalated-COFs, we investigated their electronic properties. The band structure and density of states (DOSs) of the COF-SU-00, COF-SU-01, COF-SU-02, and COF-SU-03 are computed and shown in Figure 1a, Figure 1b, Figure 1c and Figure 1d, respectively. We have plotted the band structure along a high symmetric \textit{k-}direction, $\mathrm{\Gamma-K-M-H-A-L-M-\Gamma}$, in the first Brillouin zone. All the intercalated-COFs studied here, have similar topological structure and they show similar type of electronic properties. We have found that the pristine bulk COF materials (non-intercalated-COFs) behave like an insulators as shown in Figure 1a, and when the pristine COFs are intercalated by Fe atoms, the electronic properties drastically changed. The band structures and DOSs reveal that the pristine COF i.e. COF-SU-00 has an indirect band gap around 2.60 eV resulting in an insulator, while the COF-SU-01 material gives an uncommon band structure due to the intercalation of Fe atoms (see Figure 1b). The band structure calculations show that two conduction bands just touch the Fermi energy level ($E_F$) in the $\mathrm{M-H-A-L-M}$ direction and have large electron density around $E_F$ reflected in the DOSs computation as shown in Figure 1b. The Fe atoms pushed down the conduction bands (CB) and the CB moved towards the Fermi level. From the calculated band structures and DOSs, one can see that the new proposed COF-SU-01 material is a crystal with indirect band gap, and it is about 1.20 eV as shown in Table \ref{tbl:Table1}, which resembles that of semiconductor. The indirect band gap is much smaller than the values for the pristine ordinary 3D COF materials calculated by Lukose et al.\cite{Lukose}, and the present DFT-D calculation shows that the indirect band gap is about 1.40 eV lower than the pristine COF. \\

Similarly, we have expanded our study to compute the material properties of the COF-SU-02 and COF-SU-03 materials. Both the materials show similar type behavior to the COF-SU-01 material as depicted in their band and DOSs structure calculations as shown in Figure 1c and Figure 1d, respectively. One can see clearly that the general features and characteristic peaks of the total DOSs of the COF-SU-02 and COF-SU-03 materials were changed. After adding the 4th Fe atom in the COF-SU-01, the conduction band (CB) slightly moved away from the Fermi energy level ($E_F$) and thus the CB did not touch the $E_F$. It seems that the region around the boroxine ring is insulating in the intercalated-COF, as there is no free/unpaired electron in the ring, and due to the addition of the Fe atom at the centroid of the boroxine ring, the region becomes semiconducting. It can be assumed that the $d$-sub-shell electrons of Fe are trying to interact with the electron clouds of the boroxine ring in COF-SU-02. The shape of the valence band (VB) and CB is similar type of the COF-SU-01 material. The present computation shows that the COF-SU-02 material has an indirect band gap about 1.38 eV, and the DOSs show the electron densities around the Fermi level as depicted in Figure 1c. Thus, the present study reveals that this COF-SU-02 material is a semiconductor similar to COF-SU-01. 

%\begin{figure}
%	\vspace{-20mm}
%	\includegraphics[width=0.99\linewidth]{./Figure_2}
%	\caption{Band structure and DOSs of the a) pristine COF, b) COF-SU-01, c) COF-SU-02 and d) COF-SU-03 materials.}
%	\label{BAND-DOSS_NORM}
%    \vspace{-9mm}
%\end{figure}

The shape of the CB structure was slightly changed in the COF-SU-03 due to addition of the 5th Fe atom in the COF-SU-02 material at the centroid of the triazine ring. But the shape of the VB are similar to the COF-SU-01 and COF-SU-02 materials as shown in Figure 1d. The band structure of the COF-SU-03 material shows that one conduction band is overlapped with the Fermi energy level ($E_F$) in the $\mathrm{\Gamma-K-M}$ direction making large electron density around $E_F$ which is reflected in the DOSs as shown in Figure 1d. The peaks are shown around the $E_F$ level. The $d$-subshell electrons of Fe are interacting the $p$-subshell electrons of C in the triazine ring resulting a conducting region around this ring. It seems that all the rings in the COF-SU-03 are conducting, and electron can pass through all the rings perpendicular to the basal plane in this material. In other words, due to this Fe intercalation between the two COF layers, electron can pass through one surface to the other surface of the intercalated-COFs. It also has a small indirect band gap around 1.18 eV, resulting in a semiconductor, which is well below the values for the pristine 3D COFs calculated by Lukose et al. (2.3 - 4.2 eV). \cite{Lukose} 

\begin{figure}
	\vspace{2mm}
	\includegraphics[width=0.99\linewidth]{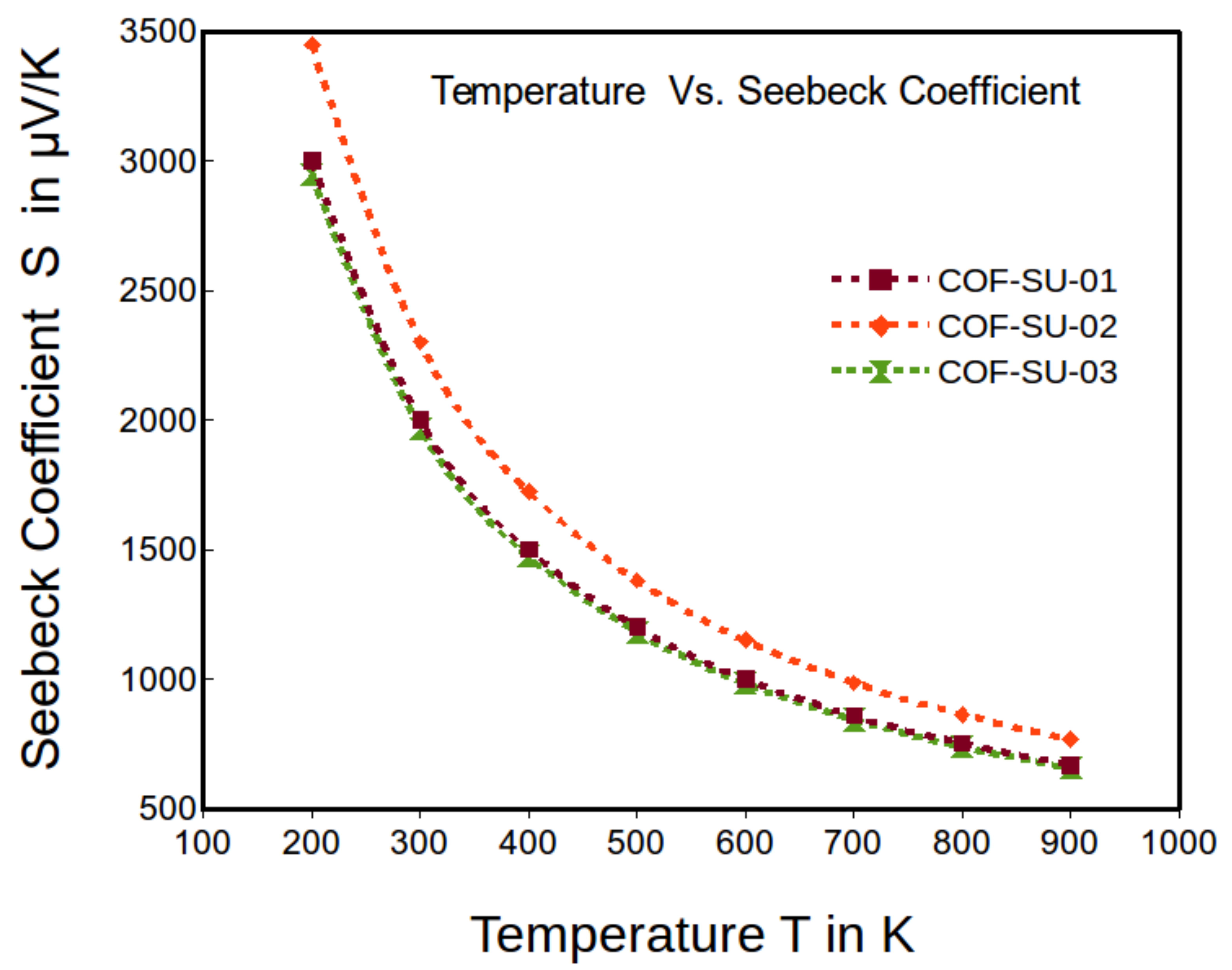}
	\caption{Temperature dependent Seebeck coefficient of the Fe intercalated COFs, COF-SU-01, COF-SU-02 and COF-SU-03 porous thermo-electric materials, estimated by the Goldsmid-Sharp theory.}
	\label{fig:Seebeck_Coefficient}
	%	\vspace{-10mm}
\end{figure}

All the thermoelectric materials are in generally semiconductors \cite{Mahan, Goldsmid}. To characterize thermoelectric materials, Seebeck coefficient calculation or measurement is generally used to estimate by computing electronic band structure properties and band gap. Goldsmid and Sharp \cite{Goldsmid} developed a mathematical expression relating the band gap and the Seebeck coefficient, $S_{max}$ (also known as thermopower), and the temperature at which it occurs ($T_{max}$). The relation between the electronic band gap $E_g$ and the Seebeck coefficient is $E_{g} = 2e|S|_{max}T_{max}$ \cite{Goldsmid}. By obtaining the DFT computed band gaps of the intercalated-COFs, COF-SU-01, COF-SU-02 and COF-SU-03, the temperature dependent Seebeck coefficient were plotted in Figure 2, which reveals that these three intercalated-COFs are thermoelectric materials. Goldsmid-Sharp method is an extremely useful tool for obtaining an estimation of Seebeck coefficient or detect the thermoelectric material, but the value of Seebeck coefficient is not an exact estimate. This result suggests that all the intercalated-COFs studied here, are thermo-electric materials making them the first crystalline porous materials.

Conceptually, to obtain a good thermoelectric material, both Seebeck coefficient and electrical conductivity must be large for the material \cite{Heremans2008}. The Seebeck coefficient is directly proportional to the density of states effective mass (${m^*}$) of the electrons in the crystal and the electrical ($\sigma$) conductivity is inversely proportional to the ${m^*}$. The band structures of the intercalated-COFs studied here, shows that these three materials should have an optimal value of the ${m^*}$ in the $M-H$ and $L-M$ region. The Seebeck coefficient can be enhanced through  a distortion of the electronic DOSs by doping or adding impurity in the system \cite{Heremans2008, Wang2011} resulting in the reduction of the thermal
conductivity. Such a situation can occur when the VB or CB of the materials resonates with the localized impurity energy level. The present computations shows the band and DOSs structures of three intercalated-COFs are distorted around the $E_F$ by adding Fe atoms in the systems, and the $E_F$ is aligned with a large and narrow peak of the DOSs resulting in enhanced Seebeck coefficient of the aforementioned materials studied here. Thus the effect of this distortion in DOSs should increase the effective mass and pronuance higher Seebeck coefficient as proposed by Synder and his co-workers  \cite{Heremans2008, Wang2011}. These three materials show the similar type of band and DOSs structures while being porous and a small indirect band gap which demonstrates a new type of porous thermoelectric intercalated-COF materials, which means that the electron can inject from valence band to conduction band at room temperature and there is no need to use extra energy to excite or promote the electrons. Our present computation also shows that the threshold adsorption wavelengths of these materials COF-SU-01, COF-SU-02, and COF-SU-03, are in the range 827 - 1035 nm, which is located in the near infrared (IR) region.   

%\begin{figure}
%    \vspace{2mm}
%	\includegraphics[width=0.99\linewidth]{./BG_Fe5_Triazene_BAND-DOSS_NORM}
%	\caption{Band structure and DOSs of the Fe5-COF material.}
%	\label{BG_Fe5_Triazene_BAND-DOSS_NORM}
%\end{figure}

In summary, the structure and electronic properties of a new kind of Fe-intercalated COF materials were studied using dispersion-corrected first principles hybrid DFT methods. The results for the new proposed intercalated-COF materials are quite interesting. Our findings suggest a new type of porous materials which have entirely different electronic properties compared to as usual porous materials such as pristine COFs or MOFs. Owing to the $\pi$-conjugated electrons around the benzene and triazine rings, the materials show flat bands in the vicinity of the $E_F$ level. The band convergence and DOSs distortion could enlarge carrier effective mass and the Seebeck coefficient, but also results in the deterioration of charge carrier mobility. The computed band structures, DOSs, and band gap demonstrate that these kind of novel intercalated-COF materials could potentially be used as semiconductors. These three intercalated-COFs can be characterized as thermoelectric porous materials. These materials have an unusual band structure making them semiconductors at room temperature compared to the typical pristine COFs which often have very large band gap around 2.6-6.0 eV. Thus this study shows that the electronic properties of the COF materials can be changed dramatically by intercalating the Fe atoms between two layers in the pristine COFs, and their band gaps can be tuned by adding Fe atoms between two layers in COFs. This approach broadens the scope and complexity of potential nanoporous COF materials. Understanding the new properties and behavior of these intercalated-COF materials will allow to design new COFs intercalated by the other transition elements in the periodic table for specific potential new applications in the near future. Further computational investigation and study will focus on the effect of the other transition metals on the electronic properties of other designed COF materials. \\

%\subsection{Supporting Information}
\textit{Supporting Information}. The optimized crystallographic information (.cif) files of the pristine and intercalated COF materials, COF-SU-00, COF-SU-01, COF-SU-02 and COF-SU-03, have been provided in the Supporting Information.

\begin{acknowledgement}

S.P. and J.L.M-C. were supported by Florida State University (FSU). J.L.M-C. gratefully acknowledges the support from the Energy and Materials Initiative at FSU, and  the support from starting funds from FSU. We are grateful to Dr. Yohanes Pramudya and Mr. Oluwagbenga Iyiola from FSU for helpful discussions and guidance with computational resources. The authors thank the High Performance Computer cluster at the Research Computing Center (RCC), FSU, for providing computational resources and support.  \\

\end{acknowledgement}

%\begin{suppinfo}

%A listing of the contents of each file supplied as Supporting Information
%should be included. For instructions on what should be included in the
%Supporting Information as well as how to prepare this material for
%publications, refer to the journal's Instructions for Authors.
%
%The following files are available free of charge.
%\begin{itemize}
%  \item Filename: brief description
%  \item Filename: brief description
%\end{itemize}

%\end{suppinfo}

%\bibliography{achemso-demo}
\bibliography{JPCL_references_2017}

\providecommand{\latin}[1]{#1}
\providecommand*\mcitethebibliography{\thebibliography}
\csname @ifundefined\endcsname{endmcitethebibliography}
  {\let\endmcitethebibliography\endthebibliography}{}
\begin{mcitethebibliography}{32}
\providecommand*\natexlab[1]{#1}
\providecommand*\mciteSetBstSublistMode[1]{}
\providecommand*\mciteSetBstMaxWidthForm[2]{}
\providecommand*\mciteBstWouldAddEndPuncttrue
  {\def\EndOfBibitem{\unskip.}}
\providecommand*\mciteBstWouldAddEndPunctfalse
  {\let\EndOfBibitem\relax}
\providecommand*\mciteSetBstMidEndSepPunct[3]{}
\providecommand*\mciteSetBstSublistLabelBeginEnd[3]{}
\providecommand*\EndOfBibitem{}
\mciteSetBstSublistMode{f}
\mciteSetBstMaxWidthForm{subitem}{(\alph{mcitesubitemcount})}
\mciteSetBstSublistLabelBeginEnd
  {\mcitemaxwidthsubitemform\space}
  {\relax}
  {\relax}

\bibitem[El-Kaderi \latin{et~al.}(2007)El-Kaderi, Hunt, Mendoza-Cort{\'{e}}s,
  C{\^{o}}t{\'{e}}, Taylor, O'Keeffe, and Yaghi]{El-Kaderi2007}
El-Kaderi,~H.~M.; Hunt,~J.~R.; Mendoza-Cort{\'{e}}s,~J.~L.;
  C{\^{o}}t{\'{e}},~A.~P.; Taylor,~R.~E.; O'Keeffe,~M.; Yaghi,~O.~M.
  \emph{Science} \textbf{2007}, \emph{316}\relax
\mciteBstWouldAddEndPuncttrue
\mciteSetBstMidEndSepPunct{\mcitedefaultmidpunct}
{\mcitedefaultendpunct}{\mcitedefaultseppunct}\relax
\EndOfBibitem
\bibitem[Cote \latin{et~al.}(2005)Cote, Benin, Ockwig, Keeffe, Matzger, and
  Yaghi]{Cote2005}
Cote,~A.~P.; Benin,~A.~I.; Ockwig,~N.~W.; Keeffe,~M.~O.; Matzger,~A.~J.;
  Yaghi,~O.~M. \emph{Science} \textbf{2005}, \emph{310}, 1166--1171\relax
\mciteBstWouldAddEndPuncttrue
\mciteSetBstMidEndSepPunct{\mcitedefaultmidpunct}
{\mcitedefaultendpunct}{\mcitedefaultseppunct}\relax
\EndOfBibitem
\bibitem[Mendoza-Cortes \latin{et~al.}(2012)Mendoza-Cortes, Goddard, Furukawa,
  and Yaghi]{Mendoza-Cortes2012}
Mendoza-Cortes,~J.~L.; Goddard,~W.~A.; Furukawa,~H.; Yaghi,~O.~M. \emph{The
  Journal of Physical Chemistry Letters} \textbf{2012}, \emph{3},
  2671--2675\relax
\mciteBstWouldAddEndPuncttrue
\mciteSetBstMidEndSepPunct{\mcitedefaultmidpunct}
{\mcitedefaultendpunct}{\mcitedefaultseppunct}\relax
\EndOfBibitem
\bibitem[Mendoza-Cortés \latin{et~al.}(2010)Mendoza-Cortés, Han, Furukawa,
  Yaghi, and Goddard]{Mendoza-Cortes2010}
Mendoza-Cortés,~J.~L.; Han,~S.~S.; Furukawa,~H.; Yaghi,~O.~M.; Goddard,~W.~A.
  \emph{The Journal of Physical Chemistry A} \textbf{2010}, \emph{114},
  10824--10833\relax
\mciteBstWouldAddEndPuncttrue
\mciteSetBstMidEndSepPunct{\mcitedefaultmidpunct}
{\mcitedefaultendpunct}{\mcitedefaultseppunct}\relax
\EndOfBibitem
\bibitem[Pramudya and Mendoza-Cortes(2016)Pramudya, and
  Mendoza-Cortes]{Pramudya}
Pramudya,~Y.; Mendoza-Cortes,~J.~L. \emph{Journal of the American Chemical
  Society} \textbf{2016}, \emph{138}, 15204--15213\relax
\mciteBstWouldAddEndPuncttrue
\mciteSetBstMidEndSepPunct{\mcitedefaultmidpunct}
{\mcitedefaultendpunct}{\mcitedefaultseppunct}\relax
\EndOfBibitem
\bibitem[Furukawa and Yaghi()Furukawa, and Yaghi]{Furukawa}
Furukawa,~H.; Yaghi,~O.~M. \relax
\mciteBstWouldAddEndPunctfalse
\mciteSetBstMidEndSepPunct{\mcitedefaultmidpunct}
{}{\mcitedefaultseppunct}\relax
\EndOfBibitem
\bibitem[{Soo Han} \latin{et~al.}(2009){Soo Han}, Mendoza-Corte, and {Goddard
  III}]{SooHan2009}
{Soo Han},~S.; Mendoza-Corte,~J.; {Goddard III},~W.~A. \emph{Chem. Soc. Rev}
  \textbf{2009}, \emph{38}, 1460--1476\relax
\mciteBstWouldAddEndPuncttrue
\mciteSetBstMidEndSepPunct{\mcitedefaultmidpunct}
{\mcitedefaultendpunct}{\mcitedefaultseppunct}\relax
\EndOfBibitem
\bibitem[Mendoza-Cort{\'{e}}s \latin{et~al.}(2012)Mendoza-Cort{\'{e}}s, Han,
  and Goddard]{Mendoza-Cortes2012a}
Mendoza-Cort{\'{e}}s,~J.~L.; Han,~S.~S.; Goddard,~W.~A. \emph{The Journal of
  Physical Chemistry A} \textbf{2012}, \emph{116}, 1621--1631\relax
\mciteBstWouldAddEndPuncttrue
\mciteSetBstMidEndSepPunct{\mcitedefaultmidpunct}
{\mcitedefaultendpunct}{\mcitedefaultseppunct}\relax
\EndOfBibitem
\bibitem[Han \latin{et~al.}(2008)Han, Furukawa, Yaghi, and Goddard]{Han2008}
Han,~S.~S.; Furukawa,~H.; Yaghi,~O.~M.; Goddard,~W.~A. \emph{Journal of the
  American Chemical Society} \textbf{2008}, \emph{130}, 11580--11581\relax
\mciteBstWouldAddEndPuncttrue
\mciteSetBstMidEndSepPunct{\mcitedefaultmidpunct}
{\mcitedefaultendpunct}{\mcitedefaultseppunct}\relax
\EndOfBibitem
\bibitem[Zhang \latin{et~al.}(2013)Zhang, Tian, Hanifi, Zhang, Sue, Zhou,
  Zhang, Zhao, Liu, and Li]{Zhang2013}
Zhang,~K.-D.; Tian,~J.; Hanifi,~D.; Zhang,~Y.; Sue,~A. C.-H.; Zhou,~T.-Y.;
  Zhang,~L.; Zhao,~X.; Liu,~Y.; Li,~Z.-T. \emph{Journal of the American
  Chemical Society} \textbf{2013}, \emph{135}, 17913--17918\relax
\mciteBstWouldAddEndPuncttrue
\mciteSetBstMidEndSepPunct{\mcitedefaultmidpunct}
{\mcitedefaultendpunct}{\mcitedefaultseppunct}\relax
\EndOfBibitem
\bibitem[Ding \latin{et~al.}(2011)Ding, Chen, Honsho, Feng, Saengsawang, Guo,
  Saeki, Seki, Irle, Nagase, Parasuk, and Jiang]{Ding2011}
Ding,~X.; Chen,~L.; Honsho,~Y.; Feng,~X.; Saengsawang,~O.; Guo,~J.; Saeki,~A.;
  Seki,~S.; Irle,~S.; Nagase,~S.; Parasuk,~V.; Jiang,~D. \emph{Journal of the
  American Chemical Society} \textbf{2011}, \emph{133}, 14510--14513\relax
\mciteBstWouldAddEndPuncttrue
\mciteSetBstMidEndSepPunct{\mcitedefaultmidpunct}
{\mcitedefaultendpunct}{\mcitedefaultseppunct}\relax
\EndOfBibitem
\bibitem[Dovesi \latin{et~al.}(2014)Dovesi, Orlando, Erba, Zicovich-Wilson,
  Civalleri, Casassa, Maschio, Ferrabone, {De La Pierre}, D'Arco, Noel, Causa,
  Rerat, and Kirtman]{Dovesi2014}
Dovesi,~R.; Orlando,~R.; Erba,~A.; Zicovich-Wilson,~C.~M.; Civalleri,~B.;
  Casassa,~S.; Maschio,~L.; Ferrabone,~M.; {De La Pierre},~M.; D'Arco,~P.;
  Noel,~Y.; Causa,~M.; Rerat,~M.; Kirtman,~B. \emph{International Journal of
  Quantum Chemistry} \textbf{2014}, \emph{114}, 1287--1317\relax
\mciteBstWouldAddEndPuncttrue
\mciteSetBstMidEndSepPunct{\mcitedefaultmidpunct}
{\mcitedefaultendpunct}{\mcitedefaultseppunct}\relax
\EndOfBibitem
\bibitem[Becke(1993)]{Becke1993}
Becke,~A.~D. \emph{The Journal of Chemical Physics} \textbf{1993}, \emph{98},
  5648--5652\relax
\mciteBstWouldAddEndPuncttrue
\mciteSetBstMidEndSepPunct{\mcitedefaultmidpunct}
{\mcitedefaultendpunct}{\mcitedefaultseppunct}\relax
\EndOfBibitem
\bibitem[Becke(1988)]{Becke1988}
Becke,~A.~D. \emph{Physical Review A} \textbf{1988}, \emph{38},
  3098--3100\relax
\mciteBstWouldAddEndPuncttrue
\mciteSetBstMidEndSepPunct{\mcitedefaultmidpunct}
{\mcitedefaultendpunct}{\mcitedefaultseppunct}\relax
\EndOfBibitem
\bibitem[Lee \latin{et~al.}(1988)Lee, Yang, and Parr]{Lee1988}
Lee,~C.; Yang,~W.; Parr,~R.~G. \emph{Physical Review B} \textbf{1988},
  \emph{37}, 785--789\relax
\mciteBstWouldAddEndPuncttrue
\mciteSetBstMidEndSepPunct{\mcitedefaultmidpunct}
{\mcitedefaultendpunct}{\mcitedefaultseppunct}\relax
\EndOfBibitem
\bibitem[Pakhira \latin{et~al.}(2015)Pakhira, Lengeling, Olatunji-Ojo,
  Caffarel, Frenklach, and Lester]{Pakhira2015}
Pakhira,~S.; Lengeling,~B.~S.; Olatunji-Ojo,~O.; Caffarel,~M.; Frenklach,~M.;
  Lester,~W.~A. \emph{The Journal of Physical Chemistry A} \textbf{2015},
  \emph{119}, 4214--4223\relax
\mciteBstWouldAddEndPuncttrue
\mciteSetBstMidEndSepPunct{\mcitedefaultmidpunct}
{\mcitedefaultendpunct}{\mcitedefaultseppunct}\relax
\EndOfBibitem
\bibitem[Pakhira \latin{et~al.}(2016)Pakhira, Singh, Olatunji-Ojo, Frenklach,
  and Lester]{Pakhira2016}
Pakhira,~S.; Singh,~R.~I.; Olatunji-Ojo,~O.; Frenklach,~M.; Lester,~W.~A.
  \emph{The Journal of Physical Chemistry A} \textbf{2016}, \emph{120},
  3602--3612\relax
\mciteBstWouldAddEndPuncttrue
\mciteSetBstMidEndSepPunct{\mcitedefaultmidpunct}
{\mcitedefaultendpunct}{\mcitedefaultseppunct}\relax
\EndOfBibitem
\bibitem[Grimme(2006)]{Grimme2006}
Grimme,~S. \emph{Journal of Computational Chemistry} \textbf{2006}, \emph{27},
  1787--1799\relax
\mciteBstWouldAddEndPuncttrue
\mciteSetBstMidEndSepPunct{\mcitedefaultmidpunct}
{\mcitedefaultendpunct}{\mcitedefaultseppunct}\relax
\EndOfBibitem
\bibitem[Pakhira \latin{et~al.}(2012)Pakhira, Sahu, Sen, and Das]{Pakhira2012}
Pakhira,~S.; Sahu,~C.; Sen,~K.; Das,~A.~K. \emph{Chemical Physics Letters}
  \textbf{2012}, \emph{549}, 6--11\relax
\mciteBstWouldAddEndPuncttrue
\mciteSetBstMidEndSepPunct{\mcitedefaultmidpunct}
{\mcitedefaultendpunct}{\mcitedefaultseppunct}\relax
\EndOfBibitem
\bibitem[Pakhira \latin{et~al.}(2013)Pakhira, Sen, Sahu, and Das]{Pakhira2013}
Pakhira,~S.; Sen,~K.; Sahu,~C.; Das,~A.~K. \emph{The Journal of Chemical
  Physics} \textbf{2013}, \emph{138}, 164319\relax
\mciteBstWouldAddEndPuncttrue
\mciteSetBstMidEndSepPunct{\mcitedefaultmidpunct}
{\mcitedefaultendpunct}{\mcitedefaultseppunct}\relax
\EndOfBibitem
\bibitem[Pakhira \latin{et~al.}(2015)Pakhira, Takayanagi, and
  Nagaoka]{Pakhira2015a}
Pakhira,~S.; Takayanagi,~M.; Nagaoka,~M. \emph{The Journal of Physical
  Chemistry C} \textbf{2015}, \emph{119}, 28789--28799\relax
\mciteBstWouldAddEndPuncttrue
\mciteSetBstMidEndSepPunct{\mcitedefaultmidpunct}
{\mcitedefaultendpunct}{\mcitedefaultseppunct}\relax
\EndOfBibitem
\bibitem[Pakhira \latin{et~al.}(2016)Pakhira, Lucht, and
  Mendoza-Cortes]{Pakhira2016b}
Pakhira,~S.; Lucht,~K.~P.; Mendoza-Cortes,~J.~L. \emph{arXiv} \textbf{2016},
  1610.04777\relax
\mciteBstWouldAddEndPuncttrue
\mciteSetBstMidEndSepPunct{\mcitedefaultmidpunct}
{\mcitedefaultendpunct}{\mcitedefaultseppunct}\relax
\EndOfBibitem
\bibitem[Lucht and Mendoza-Cortes(2015)Lucht, and Mendoza-Cortes]{Lucht2015}
Lucht,~K.~P.; Mendoza-Cortes,~J.~L. \emph{The Journal of Physical Chemistry C}
  \textbf{2015}, \emph{119}, 22838--22846\relax
\mciteBstWouldAddEndPuncttrue
\mciteSetBstMidEndSepPunct{\mcitedefaultmidpunct}
{\mcitedefaultendpunct}{\mcitedefaultseppunct}\relax
\EndOfBibitem
\bibitem[Peintinger \latin{et~al.}(2013)Peintinger, Oliveira, and
  Bredow]{Peintinger2013}
Peintinger,~M.~F.; Oliveira,~D.~V.; Bredow,~T. \emph{Journal of Computational
  Chemistry} \textbf{2013}, \emph{34}, 451--459\relax
\mciteBstWouldAddEndPuncttrue
\mciteSetBstMidEndSepPunct{\mcitedefaultmidpunct}
{\mcitedefaultendpunct}{\mcitedefaultseppunct}\relax
\EndOfBibitem
\bibitem[Monkhorst and Pack(1976)Monkhorst, and Pack]{Monkhorst1976}
Monkhorst,~H.~J.; Pack,~J.~D. \emph{Physical Review B} \textbf{1976},
  \emph{13}, 5188--5192\relax
\mciteBstWouldAddEndPuncttrue
\mciteSetBstMidEndSepPunct{\mcitedefaultmidpunct}
{\mcitedefaultendpunct}{\mcitedefaultseppunct}\relax
\EndOfBibitem
\bibitem[Momma and Izumi(2011)Momma, and Izumi]{Momma2011}
Momma,~K.; Izumi,~F. \emph{Journal of Applied Crystallography} \textbf{2011},
  \emph{44}, 1272--1276\relax
\mciteBstWouldAddEndPuncttrue
\mciteSetBstMidEndSepPunct{\mcitedefaultmidpunct}
{\mcitedefaultendpunct}{\mcitedefaultseppunct}\relax
\EndOfBibitem
\bibitem[Lukose \latin{et~al.}(2013)Lukose, Kuc, and Heine]{Lukose}
Lukose,~B.; Kuc,~A.; Heine,~T. \emph{Journal of Molecular Modeling}
  \textbf{2013}, \emph{19}, 2143--2148\relax
\mciteBstWouldAddEndPuncttrue
\mciteSetBstMidEndSepPunct{\mcitedefaultmidpunct}
{\mcitedefaultendpunct}{\mcitedefaultseppunct}\relax
\EndOfBibitem
\bibitem[Mahan and Sharp(2016)Mahan, and Sharp]{Mahan}
Mahan,~H.~J.; Sharp,~J.~W. \emph{APL Materials} \textbf{2016}, \emph{4},
  104806--1--104806--8\relax
\mciteBstWouldAddEndPuncttrue
\mciteSetBstMidEndSepPunct{\mcitedefaultmidpunct}
{\mcitedefaultendpunct}{\mcitedefaultseppunct}\relax
\EndOfBibitem
\bibitem[Goldsmid and Sharp(1999)Goldsmid, and Sharp]{Goldsmid}
Goldsmid,~H.~J.; Sharp,~J.~W. \emph{Journal of Electronic Materials}
  \textbf{1999}, \emph{28}, 869--872\relax
\mciteBstWouldAddEndPuncttrue
\mciteSetBstMidEndSepPunct{\mcitedefaultmidpunct}
{\mcitedefaultendpunct}{\mcitedefaultseppunct}\relax
\EndOfBibitem
\bibitem[Heremans \latin{et~al.}(2008)Heremans, Jovovic, Toberer, Saramat,
  Kurosaki, Charoenphakdee, Yamanaka, and Snyder]{Heremans2008}
Heremans,~J.~P.; Jovovic,~V.; Toberer,~E.~S.; Saramat,~A.; Kurosaki,~K.;
  Charoenphakdee,~A.; Yamanaka,~S.; Snyder,~G.~J. \emph{Science} \textbf{2008},
  \emph{321}, 1457--1461\relax
\mciteBstWouldAddEndPuncttrue
\mciteSetBstMidEndSepPunct{\mcitedefaultmidpunct}
{\mcitedefaultendpunct}{\mcitedefaultseppunct}\relax
\EndOfBibitem
\bibitem[Wang \latin{et~al.}(2011)Wang, Charoenphakdee, Kurosaki, Yamanaka, and
  Snyder]{Wang2011}
Wang,~H.; Charoenphakdee,~A.; Kurosaki,~K.; Yamanaka,~S.; Snyder,~G.~J.
  \emph{Physical Review B - Condensed Matter and Materials Physics}
  \textbf{2011}, \emph{83}, 1--5\relax
\mciteBstWouldAddEndPuncttrue
\mciteSetBstMidEndSepPunct{\mcitedefaultmidpunct}
{\mcitedefaultendpunct}{\mcitedefaultseppunct}\relax
\EndOfBibitem
\end{mcitethebibliography}

\clearpage
\newpage
%\section{TABLE OF CONTENTS GRAPHIC}
TABLE OF CONTENTS GRAPHIC

\begin{figure}
\centering
\includegraphics[width=0.99\linewidth]{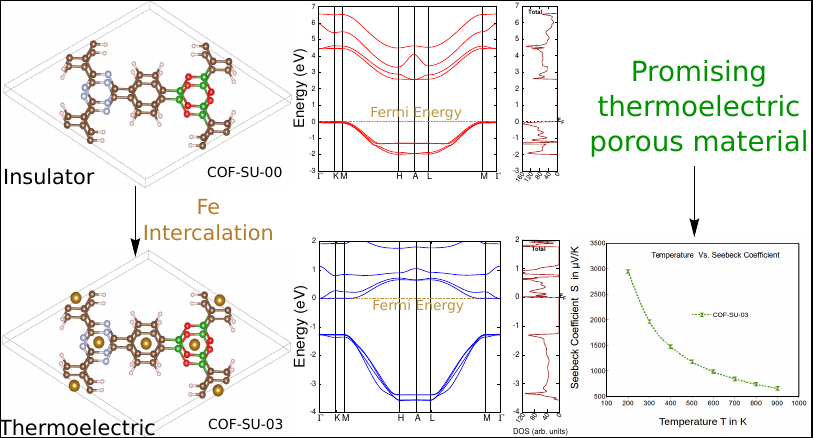}
%\caption{}
\label{fig:GOI}
\end{figure}

\end{document}